# The didactic potential of virtual information educational environment as a tool of geography students training

Olga V. Bondarenko[0000-0003-2356-2674]

Kryvyi Rih State Pedagogical University, 54, Gagarina Ave., Kryvyi Rih, 50086, Ukraine
bondarenko.olga@kdpu.edu.ua

Olena V. Pakhomova[0000-0001-5399-8116]

Oles Honchar Dnipro National University, 72, Haharina Ave., Dnipro, 49000, Ukraine
helenpah@gmail.com

Włodzimierz Lewoniewski[0000-0002-0163-5492]

Poznań University of Economics and Business, Al. Niepodleglosci 10, 61-875 Poznan, Poland
wlodzimierz.lewoniewski@ue.poznan.pl

**Abstract.** The article clarifies the concept of "virtual information educational environment" (VIEE) and examines the researchers' views on its meaning exposed in the scientific literature. The article determines the didactic potential of the virtual information educational environment for the geography students training based on the analysis of the authors' experience of blended learning by means of the Google Classroom. It also specifies the features (immersion, interactivity, and dynamism, sense of presence, continuity, and causality). The authors highlighted the advantages of virtual information educational environment implementation, such as: increase of the efficiency of the educational process by intensifying the process of cognition and interpersonal interactive communication; continuous access to multimedia content both in Google Classroom and beyond; saving student time due to the absence of necessity to work out the training material "manually"; availability of virtual pages of the virtual class; individualization of the educational process; formation of informational culture of the geography students ; and more productive learning of the educational material at the expense of IT educational facilities. Among the disadvantages the article mentions low level of computerization, insignificant quantity and low quality of software products, underestimation of the role of VIEE in the professional training of geography students, and the lack of economic stimuli, etc.

# 1 Introduction

## 1.1 The Problem Statement

The Ukrainian educational policy aims at training a competitive professional who can work effectively and efficiently in the context of rapidly changing information society. Nowadays its focus is on the transition to a virtual model of education. The geography students are required to be fluent in navigating the global information space, be able to analyze large volumes of information, be capable of life-long education [6].

However, some inconsistencies prevent the process of modernization of teacher training going smoothly. Among them, we can highlight the contradiction between contemporary social requirements and the actual preparedness of geography students to study in a virtual information educational environment (VIEE); the incompatibility of the didactic potential of information educational environment and the lack of systematic outlook on its implementation in professional teacher training.

## 1.2 Theoretical background

The analysis of scientific literature confirms the relevance of the problem under study. Thus, Mykola I. Murashko and Svitlana O. Nazarko [23] study the virtualization of the education market. Olena S. Holovnia [11] researches the systematization of the virtualization of technologies, while Irina A. Belysheva [3] studies informational and learning environment as a means of developing student cognitive autonomy. Aleksandr A. Andreev [1], Mariia A. Kyslova [16], Liubov F. Panchenko [25], Kateryna I. Slovak [26], Olha A. Obdalova [24] analyze the systematization of the virtualization of technologies. Tetiana V. Zhuravel and Nataliia I. Khaidari [38] consider the virtual educational environment as a means of formation of the student competencies. The virtual learning environment as a component of distance learning was the subject under study in Yuliia V. Falshtynska's research [9]. The advantages and disadvantages of the virtual learning environment is illuminated by Olena M. Arkhipova [2], the virtual educational environment as the innovative educational high school component is presented in the scientific researches by Maryna O. Skurativska and Serhii S. Popadiuk [31].

Some aspects of our chosen theme are considered in the studies devoted to distance education and blended learning (Myroslav I. Zhaldak [37], Volodymyr M. Kukharenko [8], Yukhym I. Mashbyts [22], Natalia V. Rashevska [26; 27], Serhiy O. Semerikov [26; 28], Andrii M. Striuk [28; 33], Yurii V. Tryus [35], Bohdan I. Shunevych [30]).

The geography study by means of the virtual educational environment and blended learning, we ground our research on, is widely presented in foreign publications. Vertual reality as an efficient way in GIS classroom teaching is studied by Jiangfan Feng [10]. The development of virtual geographic environments is researched by Fengru Huang, Hui Lin, Bin Chen [12]. Kenneth David Lynch, Bob Bednarz, James Boxall, Julie Kesby [19] work on e-learning for geography's teaching and learning spaces. Application of virtual reality in geography teaching is highlighted in works of Ivan Stojšić, Anjelija Ivkov Dzigurski, Olja Maričić, Ljubica Ivanović Bibić,

Smiljana Đukičin Vučković [32], Adem Sezer [29] considers geography teachers' usage of the internet for education purposes.

Currently the topicality of implementation of VIEE and its didactic potential for geography student training are not fully recognized in the Ukrainian scientific writings. Now there are some electronic information resources such as Google Earth, LearningApps.org, World Map Quiz, Redigo, etc. that are mostly used to study geography than to train geography students [13; 14]. In addition, some elements of the virtual educational environment, for example: e-mail, chats, forums, blogs, computer multimedia programs, electronic manuals, simulators, media resources, etc., are applied in distance and full-time studies, but due to the lack of a scientific and methodological basis, the results achieved have a fragmentary unsystematic character.

### 1.3    The objective of the article

The objective of the article is to specify the concept definition of "virtual information educational environment" and to summarize the researchers' understanding of it, and to consider its didactic potential for the training of geography students.

## 2    Results and discussion

The gradual transition of education from the knowledge to the virtual pattern reflects the process of its rapid informatization and boosts the introduction of some new terms into the scientific discourse. The comparative analysis of the meaning of the newly coined terms such as "virtual educational environment", "virtual learning environment", "information educational environment", etc., is presented in the following research (Table 1).

**Table 1.** The comparative analysis of the terms meaning

| Term | Researchers | Definition |
|---|---|---|
| Virtual education environment | Marina E. Vaindorf-Sysoeva [36] | the amount of information and means of communication of local, corporate and global computer networks, which are made and used for educational purposes by all participants in the educational process |
| | Tatiana V. Zhuravel and Natalia I. Haydari [38] | an open system enabling effective interactive self-education, based on the virtual reality technologies |
| | Maryna O. Skurativska and Serhii S. Popadiuk [31] | the organized system of informational, technological, didactic resources, various forms of computer and telecommunication interaction used by educators and students |
| Virtual learning environment | Olena M. Arkhipova [2] | a software system designed to support distance learning with an emphasis on learning, in contrast to a managed learning environment, with the emphasis on the management of the learning |

| Term | Researchers | Definition |
|---|---|---|
| | | process |
| | Yuliia V. Falshtynska [9] | a system for the learning process management created for students' learning activities, and provides the necessary replenishment and resources for successful learning and knowledge acquisition |
| Virtual learning environment / Virtual reality learning environment | Jiangfan Feng [10] | is the simulation of teaching method, thinking model, cognition manner and control means in the actual learning environment |
| Information educational environment | Olha A. Obdalova [24] | the conditions for information exchange, which are provided with the special software, aimed at satisfying the educational needs of users, usually created by cooperatively interconnected educational institutions with information exchange |
| | Liubov F. Panchenko [25] | open, nonlinear, holistic system of innovation orientation |
| Information learning environment | Irina A. Belysheva [3] | the system that reflects the interconnection of conditions and has such features as: the availability of information resources, the interactive nature of communication environment, saturation with educational resources, the possibility to change goals, methods, forms of learning organization, asynchronous use, the ability to store and accumulate information |
| | Svitlana H. Lytvynova [20] | an open system accumulating in itself deliberately created organizational-pedagogical, procedural-technological, informational resources, and with a single value-purpose basis provides innovation as a means and mechanism for the formation of components of the pedagogical culture, the formation of the professional position of teachers and the content of the forms, methods and techniques, technologies, aimed at forming a pedagogical culture of students – future teachers |
| Learning information environment | Svitlana O. Leshchuk [18] | a system of information communication and traditional means aimed at organizing and conducting a learning process focused on personal learning in the information society |
| Cloud-based learning environment | Svitlana H. Lytvynova [20] | an artificially constructed system that provides learning mobility, group collaboration between teachers and students and uses cloud technologies for effective and safe achievement of didactic goals |
| E-space / E-learning environment | Kenneth David Lynch, Bob Bednarz, James Boxall, Julie Kesby [19] | Online course materials used to support traditional campus based learning that are delivered entirely online, or provide complementary support in the form of learning materials. There are four main types of e-learning: web-supplemented, web-dependant, mixed mode, fully online. |

| Term | Researchers | Definition |
|------|-------------|------------|
| Virtual environment / Virtual geographic environment | Fengru Huang, Hui Lin, Bin Chen [12] | a concept of a virtual world that was referenced to the real world, which had five types of space, namely Internet space, data space, 3D graphical space, personal perceptual and cognitive space, and social space. |

Table 1 shows that the environment is sometimes interpreted as "information exchange", "information space". Although, most of the definitions are made in the domain of the system approach, according to which the environment is a system of interconnected, interdependent components that form a single entity performing qualitatively new function, not inherent in its separate elements. This means that the environment under study "is formed rather by educational subjects, than by technical means or electronic guides, therefore its existence is impossible beyond the communication of students, faculty, facilitators, administrators, developers of distance courses, etc." [31].

Let us consider to use the term "virtual information educational environment" in the context of geography student training.

In such an environment, the nature of the interaction of all participants of learning (a student, a student group, a teacher) is fundamentally changing. We deliberately do not use the term "study" for we interpret it as a cognitive activity of students who under the guidance of a teacher master knowledge, skills, and develop cognitive abilities. The involvement of the geography students in the information educational environment means that a teacher changes his role of a head and mentor to a tutor, facilitator, and moderator. Teacher strives to help a student find an individual educational route (an individual way of personal potential realization as a student that is made taking into account his abilities, interests, needs, motivation, opportunities and experience).

Geography, socio-economic geography in particular, is a discipline with dynamic content. So the knowledge received by former student at the higher school in the past can become "out of date for one day" in relation to the present one. That is why it is of great importance to teach a student to collect and analyze the necessary information by his/her own using available statistical sources such as countrymeters.info, ukrstat.gov.ua; to process the records and statistics of international organizations such as UN, WHO, etc.; to use information in the form of self-created cartographic works, schemes, drawings based on DataGraf, QuickMap, and Google Earth; to work with interactive maps such as MigrationsMap, kartograph.org, Setera.org, World Map Quiz, Redigo, Mapillary, etc.

Modern students live in a media environment where the use of computers, Internet resources and mobile devices is the part of their everyday life. They are, according to Aleksander Kuleshov "digitally born", and this fact cannot be ignored. Therefore, "key properties" [15] VIEE, which can realize its didactic potential to some extent, should be taken into account while training of geography students. They are in particular:

— Immersion, the ability to be an active doer instead of a passive viewer (for example, not to state and percept the volume of world's population, but try on a role of an expert UNFPA, who is capable of predicting the dynamics of the population over the coming decades);

— Interactivity, the active interaction of education process participants among each other and with an artificial environment (for example, to be able to carry on a peer survey in order to find out the family and childbirth plans). Interpretation of the results implies finding out the nature of demographic behavior of the population in countries and regions with different types of reproduction of the population);

— Dynamism, variability, transience of events (for example, the skill of comparative analysis of content and processes for the creation of medieval and modern digital geographic maps);

— Sense of presence (for example, to be able to role-play the characteristics of demographic behavior of representatives of various social strata of Africa);

— Continuity, the ability for continuous interaction of participants in the educational process (off-line, on-line, etc.);

— Causality, the ability to identify the causal relationships among physical and geographical, socio-economic, cultural-historical phenomena and processes, and to visualize them with multimedia.

Moreover, global informatization of society has caused the problem of cognitive overload (when the number of operations that human brain must perform exceeds its capacity) [21]. This problem concerns geography students as well, because their training involves the memorization of a significant amount of factual material (geographical nomenclature, quantitative and qualitative parameters of the population of the world and regions, indicators of socio-economic development of countries, algorithms of work with geodetic instruments, trends and patterns of physical development geographic, socio-economic, cultural-historical phenomena and processes, etc.). On the other hand, the VIEE enables a student to visualize information, choose the form and rate of education, the level of complexity of the task, methods and means of training. Its creation encourages a student to make more personalized educational and practice-oriented objectives. Aleksei N. Leontiev emphasize the importance of personal sense of learning objectives and identify the later as the reflection of the relation of an activity motive to the purpose of an action in the mind of an individual [17].

As the scientific works do not share the commonly accepted definition of the term "virtual information educational environment", there is an urgent need to define our own one taking into account the definitions given by other authors, and the specifics of teacher work and the requirements for the personality of the geography teacher. Basing on the researches of Maryna O. Skurativska and Serhii S. Popadiuk [31], we define the "virtual information educational environment" as a holistic, organized system of various resources (informational, didactic, technological) and forms of interaction of teachers and students (synchronous, asynchronous, full-time, remote, computer and telecommunication), aimed at creating student individual educational trajectory.

Any platforms as commercial as free suitable for blended learning such as Blackboard, Bodington, CloudSchool, Edmodo, Google Classroom, Moodle, etc. can serve as the basis for creating a virtual information educational environment.

The authors of this article have the experience in scaffolding of blended learning of geography students by using Google Classroom [4; 5; 7], Fig. 1.

It should be noted that the educational process organized in the VIEE, based on the Google Classroom, is similar in its structure to the traditional one and embraces the objective-motivational, content-operational, emotional-volitional, evaluative-productive components. Although, all of them are fulfilled with the specific technologies and tools (IT communication tools and IT educational facilities).

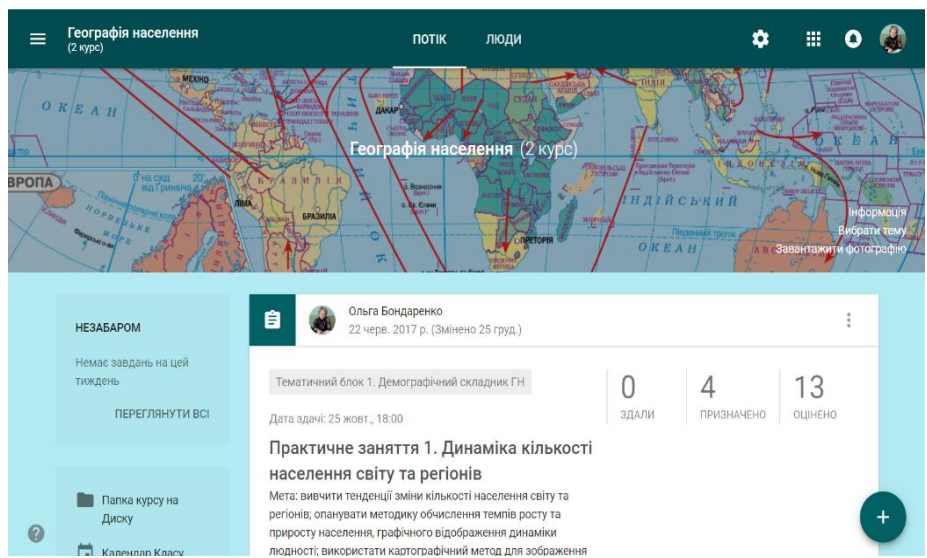

**Fig. 1.** The main page view of the course "Population Geography"

## 3 Conclusion

1. To summarize the stated above we may claim the advantages of VIEE:

— increase of the efficiency of the educational process by intensifying the process of cognition and interpersonal interactive communication [23; 34];
— continuous access to multimedia content both in Google Classroom and beyond;
— saving student time due to the absence of necessity to work out the training material "manually";
— availability of virtual pages of the virtual class;
— individualization of the educational process;
— formation of informational culture of the geography students;
— more productive learning of the educational material at the expense of IT educational facilities.

2. On the basis of the analysis of scientific literature (Olena M. Arkhipova [2], Yuliia V. Falshtynska [9], Mykola I. Murashko and Svitlana O. Nazarko [23], and own pedagogical experience we can state that despite the obvious number of advantages of VIEE, its adoption among the students and teaching community is slow in Ukraine. To a certain extent, this situation has arisen because of the low level of computerization of institutions of higher education and the lack of economic stimuli. Depending on the specifics of the discipline, the quantity and quality of the necessary software products can be quite low. The underestimation of the role of VIEE in the professional formation of geography students also prevents its adoption to certain extend.
3. We anticipate the further study of the problem in the development of the systems of electronic courses on Google Classroom platforms and Moodle for the organization of distance learning of the geography students of the correspondence department and people with special educational needs.

ekonomichnoi i sotsialnoi heohrafii svitu (Using Google Classroom towards a study of the regional economic and social geography of the world). In: Implementation of ICT in the educational process of educational institutions, pp. 3–5. Poltava V.G. Korolenko National Pedagogical University, Poltava (2016)